# Machine Learning Approach for Transforming Scattering Parameters to Complex Permittivity

Authors: Robert Tempke[*, a], Liam Thomas[*], Christina Wildfire[b,c], Dushyant Shekhawat[c], Terence Musho[*,c,z]

[*]Department of Mechanical & Aerospace Engineering, West Virginia University, Morgantown, WV, USA.
[a]Oak Ridge Institute of Science and Education (ORISE).
[b]Leidos Research Support Team (LRST), Morgantown, WV, USA.
[c]National Energy Technology Laboratory – US Department of Energy, Morgantown, WV, USA.
[z]Corresponding Author: tdmusho@mail.wvu.edu

Abstract:

This study investigates the application of an artificial neural network to predict the complex dielectric properties of granular catalysts commonly used in microwave reaction chemistry. The study utilizes finite element electromagnetic simulations and two-dimensional convolutional neural networks to solve for a large solution space of varying dielectrics. This convolutional neural network was trained using a supervised learning approach and a common backpropagation. The frequency range of interest was between 0.1 – 13.5 GHz with the real part of the dielectric constants ranging from 1 – 100 and the imaginary part ranging from 0 – 0.2. The network was double validated using experimental data collected from a coaxial airline. The model was demonstrated to convert either experimental or computational derived scattering parameter to complex permittivities. Moreover, the model eliminates the need for iterative solutions that often have difficulty with the piecewise continuous nature of frequency dependent scattering parameters.

Introduction

With recent advanced in dielectrics and their synthesis techniques, there has been a need to increase the fidelity of inverse models to predict the dielectric properties of materials based on a measurable observable. In most cases and the case in this study, these observables are scattering parameters derived from coaxial transmission line testing of the dielectrics. Having confidence in the measurement of the dielectric is important for fundamental device design in the fields of microwave engineering, microwave material processing, and microwave chemistry. It is hypothesized that new advancements in machine learning could allow for a model trained on a wide combination of synthetically generated solutions that encompass the entire range of feasible values and frequencies, which could be applied to experimental data to produce stable, fast, and accurate results.

At GHz frequencies, the electromagnetic (EM) interactions are quantized by a material's dielectric properties or the dynamics of dipole interactions. The dielectric constant of a material is the ability of the material to store electrical energy. While the loss tangent of a material is a quantification of how well it will be able to transform that energy into heat. For microwave reactions, the dielectric properties of a material are one of the most informative metrics for how well the material will heat in a microwave. These properties can be hard to characterize as the polarization of particles changes even among identical products [1]-[2]. These dielectric properties are being explored for a wide range of applications including microwave chemical reactions. To date, the characterization of these properties has been done using a multitude of inverse mathematical techniques. Depending on which technique is employed there will always be discrepancies between solutions. Many of the interactive techniques require initial guesses to avoid discontinuities arising from the resonance of the system [3]–[5]. Recent computational advancements in the ability to conduct large numbers of permutations of solutions with high accuracy have ushered in the potential to revisit many of these inverse methods using a machine learning approach.

The complexity of many dielectric materials such as in the case in microwave chemistry and the use of heterogeneous (multi-component, macroscopic, granular) catalysts leads to inaccuracies in dielectric constant calculations, steaming from non-standard synthesis procedural approach and therefore high dimensionality of inverse space. The most common method of determining complex dielectric properties are from calculations based on scattering parameters (S-parameters). Multiple measurement techniques utilize S-parameters, such as a rectangular free-space waveguide, open-ended probe, free space, resonant cavity, parallel plate and coaxial precision airline [1], [6]–[8]. These different methods utilize different inverse techniques such as Nicholson-Ross-Weir (NRW), NRW polynomial, NIST Iterative, NIST non-iterative, and the short circuit line (SCL) methods [3], [5], [9].

Artificial neural networks (ANNs) and convolutional neural networks (CNN) are a subset of machine learning that lend themselves well to material science problems. The usage of these networks has been steadily on the rise over the past decade, with more and more studies investigating the possibilities of ANNs to map non-linear relationships [13]. ANNs are part of the biologically inspired computational techniques used in different artificial intelligence applications [14], [15]. ANNs have been used in many applications of chemistry, material science, and microwave engineering [15]. Machine learning models are commonly accepted as more accurate than traditional linear and non-linear statistical regression methods when dealing with highly dimensional inputs [24]. This advantage of machine learning algorithms only increases as the dimensionality and non-linearity of the relationships increases [25]–[27].

Typically, machine learning models and ANNs, in particular, are used in literature to try and relate complex geometric parameters or certain material fingerprints to dielectric constants [14], [16]. However, Tuck and Coad [17] showed that ANNs can be used to calculate the dielectric properties of liquids directly from the S-parameters using a coaxial probe method without the need for de-embedding the data first. By calculating the dielectric properties directly from the recorded S-parameters without needing to de-embed the data in the time domain Tuck and Coad were able to achieve a significant reduction in the intrinsic error. This was because the ANN was able to capture the realities of the non-ideal system by training the ANN on vectors of reflected coefficient data and correlating that to the permittivities of the substance being studied. This method avoided the need for any parametric models of the cable such as those used by Stuchly et al. [18].

Chen et al. [19] demonstrated that this same de-embedding approach was suited for different probe geometries. More importantly, Chen et al. were able to show that finite different time-domain (FDTD) simulation data can be used to accurately train an ANN for prediction on experimental data. They even postulate that this method would work for powdery materials and at high temperatures. These ANNs however, were limited in scope to only liquids and by only learning the non-linear and complex relationships between the reflected coefficient of the scattering parameters (S11) and the dielectric properties rather than in more advanced measurement techniques in which S11 and S22 are utilized. Regardless the results were fast and extremely accurate calculations of dielectric properties of a combination of different liquids.

This study investigates and implements a machine learning algorithm for the use of calculating the dielectric properties of solid materials in a coaxial airline.  This approach like Tuck and Coad's will allow for the calculation of the dielectric material to take place directly from the measured S-parameters without the need to de-embed the air in the coaxial airline. Thus, simplifying the mathematics and intrinsic error. The approach is to use supervised learning to teach and validate an algorithm using simulation data. Then to test the CNN model on experimental scattering parameters collected for known dielectric materials. This approach was selected because of the precise control of the input and outputs being used in training the system and its reproducibility. The system looks to achieve increased accuracy over previous models by utilizing the additional input parameters available when testing in the coaxial airline. The previous studies were only able to utilize the reflected coefficient of the S-parameters (S11) because of the limitations of the coaxial probe method. The coaxial airline method provides both the forward and reflected coefficients of the S-parameters (S11 and S21).

The approach for this study is to generate simulation solutions for a wide variety of dielectrics. In addition to varying the dielectric properties in the simulation, the specimen length within the coaxial line was also varied. It is postulated that for any sample length and dielectric constant there exists a unique set of inputs (S-parameters) that generate that solution. By teaching a machine learning algorithm these relationships based on multiple conversion methods a more robust and accurate solution can be obtained than previously existed. Utilizing the simulation results with machine learning can potentially result in a much faster and less computationally intensive solution methodology. Together these techniques can provide a new solution method for converting S-parameters to dielectric properties.

1. Classic Measurement Methods

The most commonly used conversion model for S-parameters to dielectric properties is the NRW method as it gives information on both the electric and magnetic properties of a material. The NRW provides a direct calculation of the dielectric properties from the S-parameters. This method utilizes the $S_{11}$ and $S_{21}$ parameters which are the ratio of the wave reflected from the material and the wave that passes through the material respectively [3]. It is a very robust method that can solve for many different types of materials. The popular Keysight vector network analyzer utilizes this approach as the standard option when calculation dielectric properties. However, the main drawback to this method is that the solution diverges at frequencies of integer multiples of ½ wavelength for low loss materials. This leads to this method performing better with shorter samples [5], [20].

Another popular method used in literature is the NRW polynomial method, this method takes the NRW conversion method and fits a polynomial to the dielectric properties. This eliminates the discontinuity peaks at ½ wavelength but turns the entire solution into a close approximation rather than a high precision measurement[9]. The NIST iterative conversation method is another method for calculating the dielectric properties that utilize the Newton-Raphson's root finding method to calculate the dielectric properties. This method avoids the discontinuities that happen when using the NRW method by requiring a good initial guess and is good for long samples and low loss materials [3]. Without a good initial estimate, the solution will diverge and/or be highly inaccurate. The NIST iterative method also assumes that permeability is equal to unity making it applicable for only nonmagnetic materials [6].

The other NIST method is a non-iterative method that closely resembles the NRW method but assumes the permeability is equal to unity. Unlike many of the other methods, there are no sample length criteria for this method any arbitrary sample length is acceptable [3]. The biggest drawback to this method is the smaller scope of materials it can measure because of the non-magnetic assumptions [6]. A unique measurement method among transmission line measurements of S-parameters is the SCL method.

Calculations are performed using only the $S_{11}$ parameter and accurate sample positional information to calculate the complex dielectric properties of a material. The SCL method uses Newton-Raphson's numeric approach to calculate dielectric properties. The simplicity of the inputs for this method makes it suited for broadband measurements and long samples with low loss. Like all the methods but the NRW methods the SCL method also assumes a permeability equal to one [3].

2. Computational Details
    2.1 Material Dataset Generation

The calculations of S-parameters for varying dielectric properties were determined using a finite element (FE) EM wave modeling software COMSOL Multiphysics® [21]. All solutions were solved in the frequency domain, using the finite element frequency domain (FEFD) approach. While it is beyond the scope of this study to elaborate on the advantages of FEFD method over an FDTD method in detail, the advantage is twofold. First, the FE approach is an implicit method that relies on a minimization method while the FD involves a stability criterion dependent on the mesh characteristics. Secondly, by solving in the frequency domain the computational time is significantly reduced by eliminating a time stepping criteria. While these advantages do not apply to all problems, especially large (time and spatial) non-linear problems, the frequency domain was an appropriate for the following 2D axis-symmetric linear (steady state, non-temperature dependent properties) study.

Using the FEFD model a series of parametric sweeps of both the real and imaginary portions of the dielectric properties was performed to encompass all naturally occurring dielectric materials. Properties were swept from a real dielectric constant of 1 to 100 in increments of 0.5 while the imaginary portion of the dielectric properties was varied in from 0 to 0.2 in increments of 0.05. While the length of the sample was increased from 10 mm to 50 mm in 1 mm increments. These dielectric properties were swept in correspondence with a frequency range of 0.1 to 13.5 GHz at 51 equally spaced points. The computational model was set up to represent a two-port vector network analyzer with a high precision coaxial airline of length 150 mm. The airline is modelled based on the experimental airline used for validation. The coaxial airline is a HP model no. 85051-60010 with a 0.70 cm diameter.

A 2D axis-symmetric FE model was constructed, which represented the 150mm coaxial airline that is used in the experimental measurement setup. Figure 1 is an illustration of the coaxial airline modeled with the FE solver software. The walls of the airline, as well as the center electrode, were assumed to be perfect electrical conductors. Due to the axis-symmetric assumption it is assumed only a 2D representation of a slice in the +r and +z directions needed to be constructed. The plane was partitioned at z=10 mm to form to material regions increasing after each full parametric sweep. The region between z=0 mm and z=10-50 mm will be defined as the sample region. The remaining will be assigned air (vacuum, $\varepsilon_r=\mu_r=1$). Two ports were defined at extremes in the z-dir. Port 1 was defined at z=150 mm and Port 2 was defined at z=0 mm. A coaxial boundary condition (TEM mode) was specified for both ports. The scattering parameter was measured at both port planes in the absence of de-embedding as would be representative of experimental measurement where de-embedding has taken place during the calculation of the dielectric properties. Figure 1B is a contour plot of the radial electric field (Er) at 13 GHz and 1W of input power at Port 1. For the remainder of the study, 0.1W will be used as the input power at both ports with the understanding that 1) material properties are linear and are not changed by field strength or temperature, 2) scattering parameters are a function of normalized power, and 3) the experimental network analyzer will utilize much lower port power. The color contours of Figure 1B and the inset image confirm the radial electric field synonymous with a TEM mode.

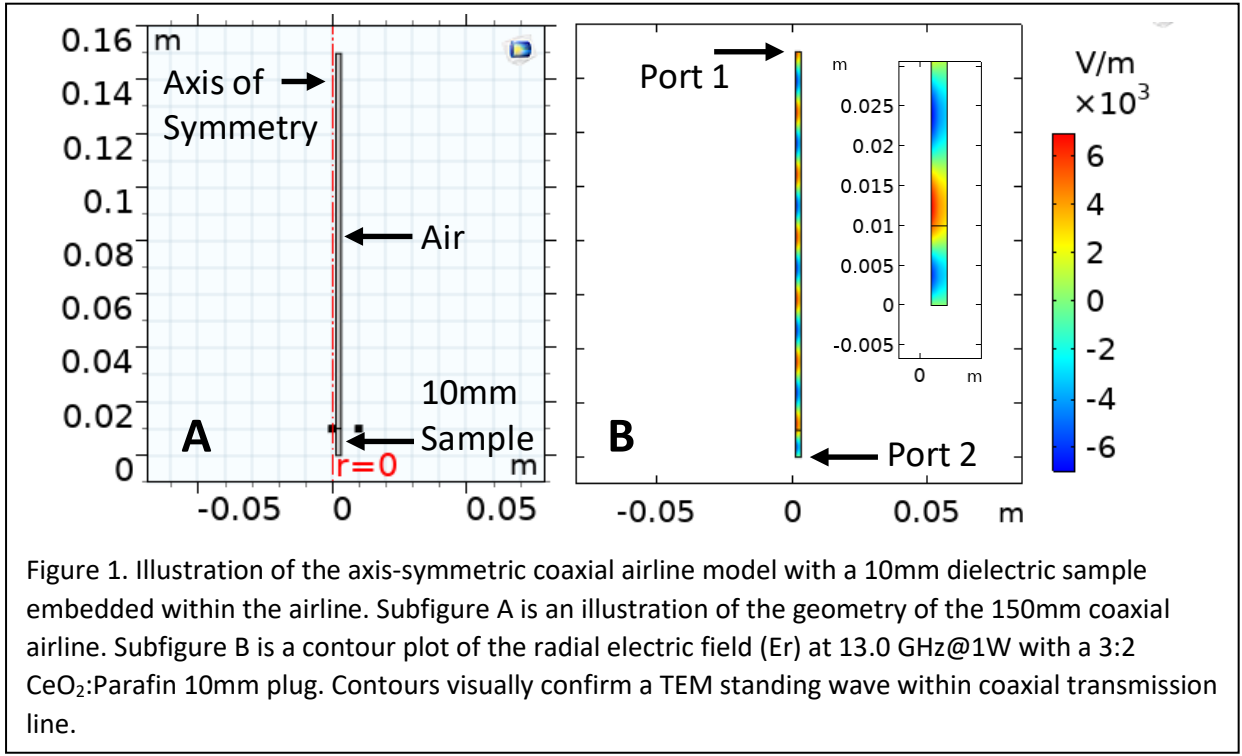

Figure 1. Illustration of the axis-symmetric coaxial airline model with a 10mm dielectric sample embedded within the airline. Subfigure A is an illustration of the geometry of the 150mm coaxial airline. Subfigure B is a contour plot of the radial electric field (Er) at 13.0 GHz@1W with a 3:2 $CeO_2$:Parafin 10mm plug. Contours visually confirm a TEM standing wave within coaxial transmission line.

2.2 Artificial Neural Network Implementation

The ANN was developed using open-source TensorFlow developed by Google for ease of implementation with all data scaled to be within the same power factor. The ANN used two different approaches one in which all 5 input features of, frequency, the magnitude of S11 and S21, and the phase in radians of S11 and S21 were studied independently of one another. This approach would allow for researchers to get discreate answers at any frequency point independent of the solutions to previous frequencies. The other approach was to look at all the inputs for a given dielectric at once, in this case, all 51 data points from 0.1 to 13.5 GHz. To attempt to pull the latent information that exists in the transition between wavelengths. The data was broken down into 3 different sets 60% was allocated to training data, 20% to validation data and 20% to test data. With the experimental data being kept separate until a suitable algorithm had been created. This breakdown allows the algorithm to be tested on unseen data ensuring that it was not overfitted to the training and validation data before it was tested on experimental data. To achieve the ideal performance of this network multiple different loss functions were looked at as well as different combinations of the number of neurons and number of hidden layers. Different regularizes were investigated to help encourage convergence. These different combinations were evaluated using the mean squared error (MSE) and the mean absolute error (MAE).

To ensure optimization of these parameters the ANN utilized the ReLU activation function and the Adadelta optimizer. The network also introduced gaussian noise into the training data to represent real world errors in experimental setups. The ReLU activation function was chosen because of its proven ability to represent sparsity [22], [23]. Sparsity is useful in ANN because of its ability to imitate a biological neural network. Sparsity in an ANN allows for models to have better predictive power with less noise and overfitting by encouraging neurons to only process meaningful aspects of the problem [23], [24]. In the work by Maas et al. [25] and Narang et al. [26], they demonstrated that increased sparsity helped to improve an ANNs training performance and reduce computational time for several problems.

The Adadelta optimizer is a gradient descent method that utilizes a dynamic updating system. The system adapts using first-order information and stochastic gradient descent which reduces its computational cost over many of the other optimizers available [27]. One of the key advantages of this optimizer for the ANN system of interest is that it requires no human training of the learning rate and can handle training data that may have lower signal to noise ratios. These two hyperparameters were chosen after some initial data testing and held constant for the remainder of the study. The first scheme when considering frequency points independently, two convolutional layers and two fully connected layers. The second scheme in which an array of frequency points is used, two convolutional layers, one max pooling layer and two fully connected layers were used.

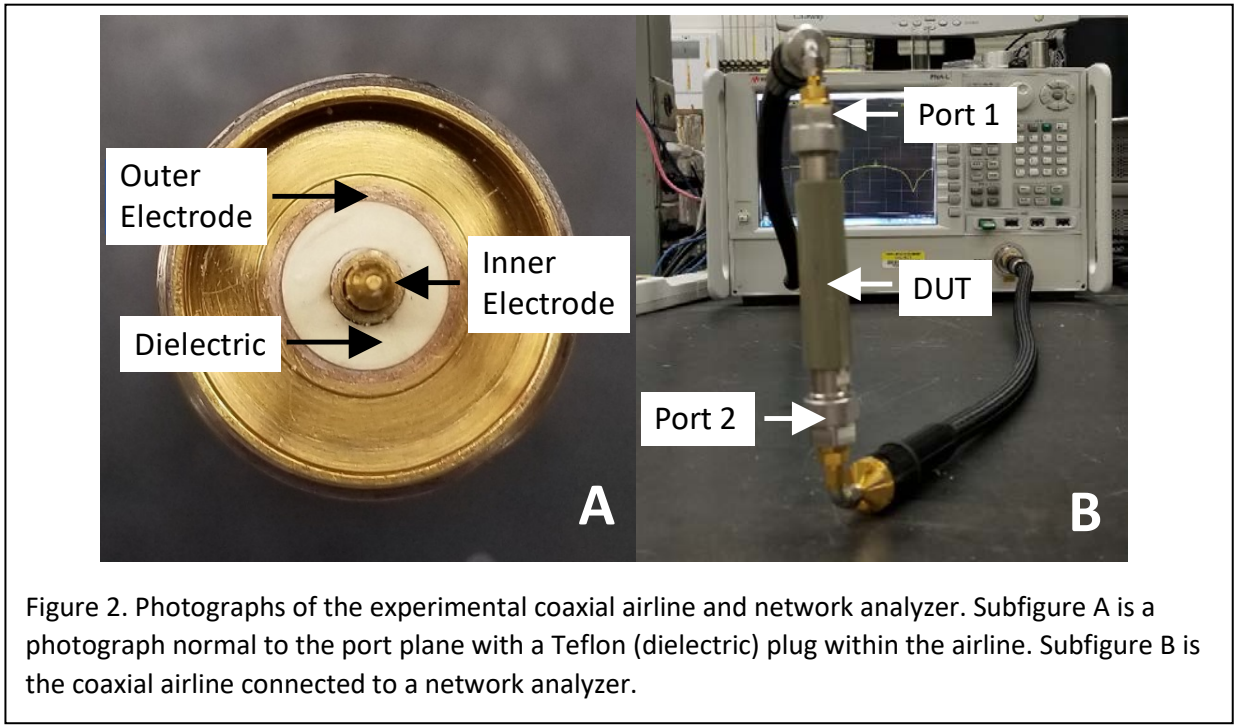

Figure 2. Photographs of the experimental coaxial airline and network analyzer. Subfigure A is a photograph normal to the port plane with a Teflon (dielectric) plug within the airline. Subfigure B is the coaxial airline connected to a network analyzer.

2.3 Experimental Data Collection Method

Data was collected on a Teflon (PTFE) plug, this replicated the standard NRW test that validated their measurement model. To take a measurement the sample was loaded into the airline with the center electrode in place, as shown in Figure 2A. All interfaces between the airline and cables were thoroughly cleaned using isopropyl alcohol and dried using dry compressed air. Each test was conducted with a frequency range from 0.1 to 13.5 GHz. The scattering parameters were recorded at 51 equally spaced points within this range. The relative dielectric constant for each point was using the NRW method. All measurements reported in the study were conducted using a 0.70 cm diameter coaxial airline (HP model no. 85051-60010), as shown in Figure 2A, and connected to a Keysight N5231A PNA-L microwave network analyzer shown in Figure 2B.

3. Results and Discussion
    3.1 Correlation Analysis Method

The simulation derived dielectric datasets consisted of 330,813 values with real portions of the dielectric constant (ε') ranging from 1 to 100 and the imaginary portion of the dielectric constant (ε") ranging from 0 to 0.2. The corresponding inputs of S-parameters include the magnitude and phase of S11 and S21. Because the system is symmetric S11=S22 and S21=S12 and therefore only S11 and S21 are necessary. To create an efficient and accurate machine learning model a statistical analysis of the input features needed to be performed to determine their significance on output targets. Strong correlations can be both good and bad for ANNs, strong correlations can help to reduce the number of input features needed for the network. They can also skew the network towards harmful bias creating multicollinearity with a single input and the target feature. Which can result in small changes to the input data leading to large changes in the model. To check on these traits a Pearson correlation was performed between all the input features and the output targets [28]. Table 1 is a summary of the correlation between the inputs and outputs. With a 1.0 meaning a very strong positive correlation and a -1.0 corresponding to an inversely related correlation. The complex dielectric is defined as $\varepsilon_r = \varepsilon' - i\varepsilon''$, where ε' is the real portion and ε" is the imaginary portion. The scattering parameter magnitude is denoted as |S11| and |S21|. The associated phase angle of the scattering parameter is denoted as ∠S11 and ∠S21.

|              | |S11| | |S21| | ∠S11 | ∠S21 | ε'  | ε"  |
|--------------|-------|-------|------|------|-----|-----|
| |S11| (Input) | 1.0   | ---   | ---  | ---  | --- | --- |
| |S21| (Input) | -0.9  | 1.0   | ---  | ---  | --- | --- |
| ∠S11 (Input)  | 0.0   | 0.0   | 1.0  | ---  | --- | --- |
| ∠S21 (Input)  | 0.0   | 0.0   | 0.0  | 1.0  | --- | --- |
| ε' (Output)   | 0.8   | -0.9  | 0.0  | 0.0  | 1.0 | --- |
| ε" (Output)   | 0.0   | -0.2  | 0.0  | 0.0  | 0.0 | 1.0 |

Table 1. Correlation matrix for input and output parameters. Values range from -1 to 1. Negative values are associated with inverse correlation.

As seen in Table 1 the magnitudes of S11 and S21 are strongly correlated to the real part of the dielectric constant. The features that correlate linearly to a dielectric constant are the magnitude of the wave that is reflected from a material and magnitude of the same wave that passes through the material. This correlation of magnitudes is expected since the real part of the dielectric constant is the ability of a material to store energy. It is noted from Table 1 that there is no linear correlation between the phase angle and the magnitude of scattering parameters. There is also a lack of correlation between the phase angle and the dielectric constant. While this is correct for the assumed linear materials and constant sample geometry (plug length) some caution must be taken with this correlation. If the phase angles of the scattering parameters were eliminated it would make this system highly linearly correlated to the magnitude of the scattering parameters. This would result in more system performance as small changes in the magnitude of the scattering parameters would have large effects on the output of the system (dielectric properties). CNN's are uniquely suited to this type of problem because not only do they perform their calculations in high dimensionality, but they use convolutional math applied over the input data. Therefore, the inclusion of the phase angle allows the network to eliminate its dependence on the magnitude of the scattering parameters as seen in other multi-layer perceptron networks.

To achieve a more extensive understanding of the different relationships between input features and output targets joint plots were created for each input. These are shown in Figure 3A-D where the entire spectrum of dielectric properties as a function of the inputs. The darker regions of the contour plot representing a stronger correlation. From these plots, a better understanding of the correlation coefficient from Table 1 can be gained. The strong positive and negative correlation for the magnitudes of S11 and S21 can be seen in Figures 3a and 3b. However, the figure illustrates that there is a direct effect on correlation based on the magnitude of the dielectric properties (|S11| and |S21|). Lower dielectric constants ($\varepsilon'<20$) express little to no correlation between the input parameters and the output. While high dielectric constants (>40) show an increasingly strong correlation between the inputs and the output as the dielectric constant approaches 100. This growing correlation will provide a unique challenge to the design of the ANN architecture as traditional approaches to strong and weak correlation architecture will be insufficient to capture the unique relationship.

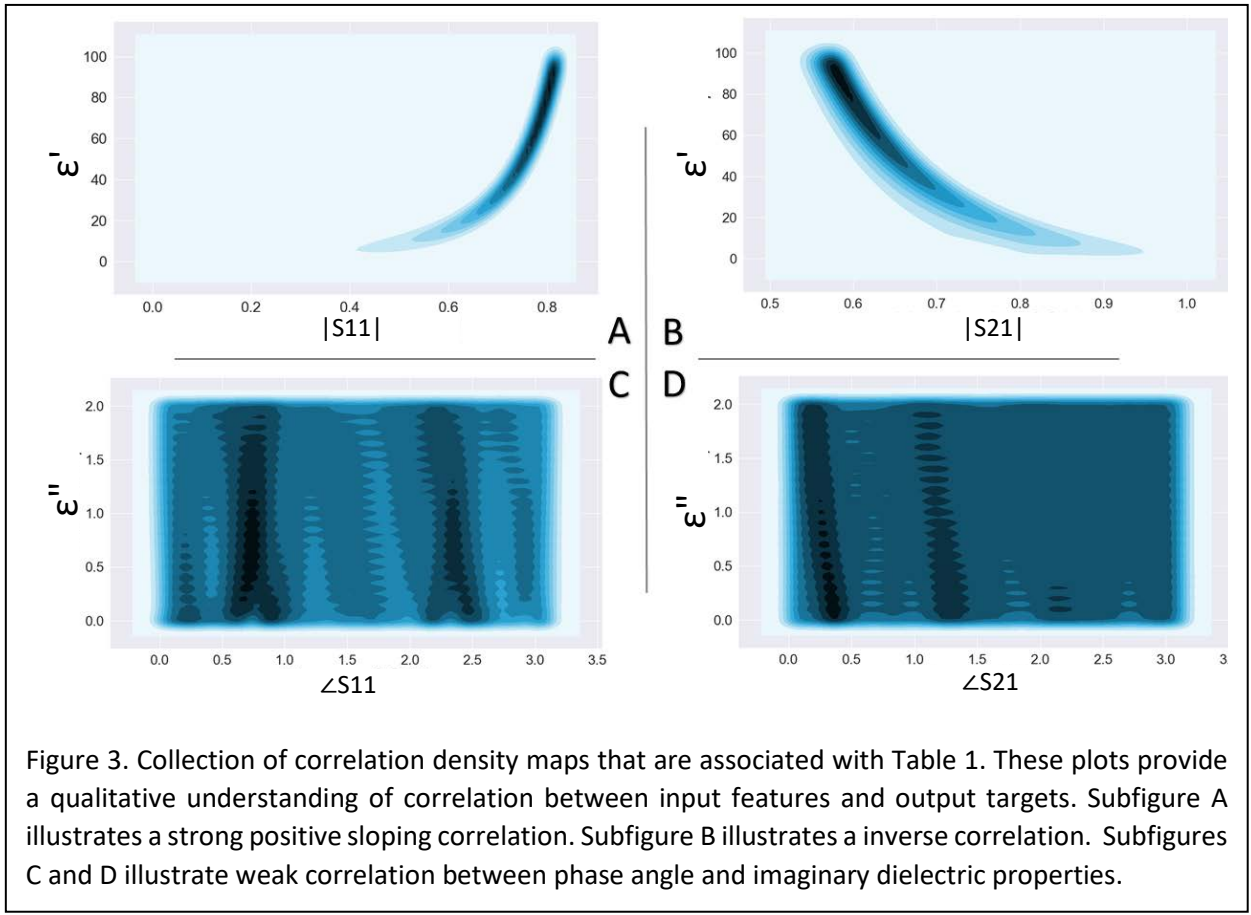

Figure 3. Collection of correlation density maps that are associated with Table 1. These plots provide a qualitative understanding of correlation between input features and output targets. Subfigure A illustrates a strong positive sloping correlation. Subfigure B illustrates a inverse correlation. Subfigures C and D illustrate weak correlation between phase angle and imaginary dielectric properties.

4. Results
    6.1 ANN Results

The trained neural network was used to predict on randomly generated test data that the ANN was not explicitly trained or validated on. Multiple models with a varying number of convolutional layers, hidden layers, neurons, and loss metrics were evaluated for their applicability in the calculation of the complex dielectric properties. Each test was run for 500 epochs to allow for convergence to an optimized set of weights and used the relu activation function along with the adadelta optimizer. The training set used in the network had a mean dielectric constant of 50.4 and a standard deviation of 28.6 while the test set had a mean of 50.8 and a standard deviation of 28.7. This similarity confirms that the test datasets contains a good representation of the whole dataset. Demonstrating that the data was well randomized and ANN performance was not due to the selection of a certain sub-dataset.

Two different approaches were used to study the validity of this approach, the first was to train the network for each frequency point of each dielectric independently, the other approach was to train the entire frequency spectrum (0.1 to 13.5 GHz) as a whole. The latter approach permitted the algorithm to learn interpoint traits.

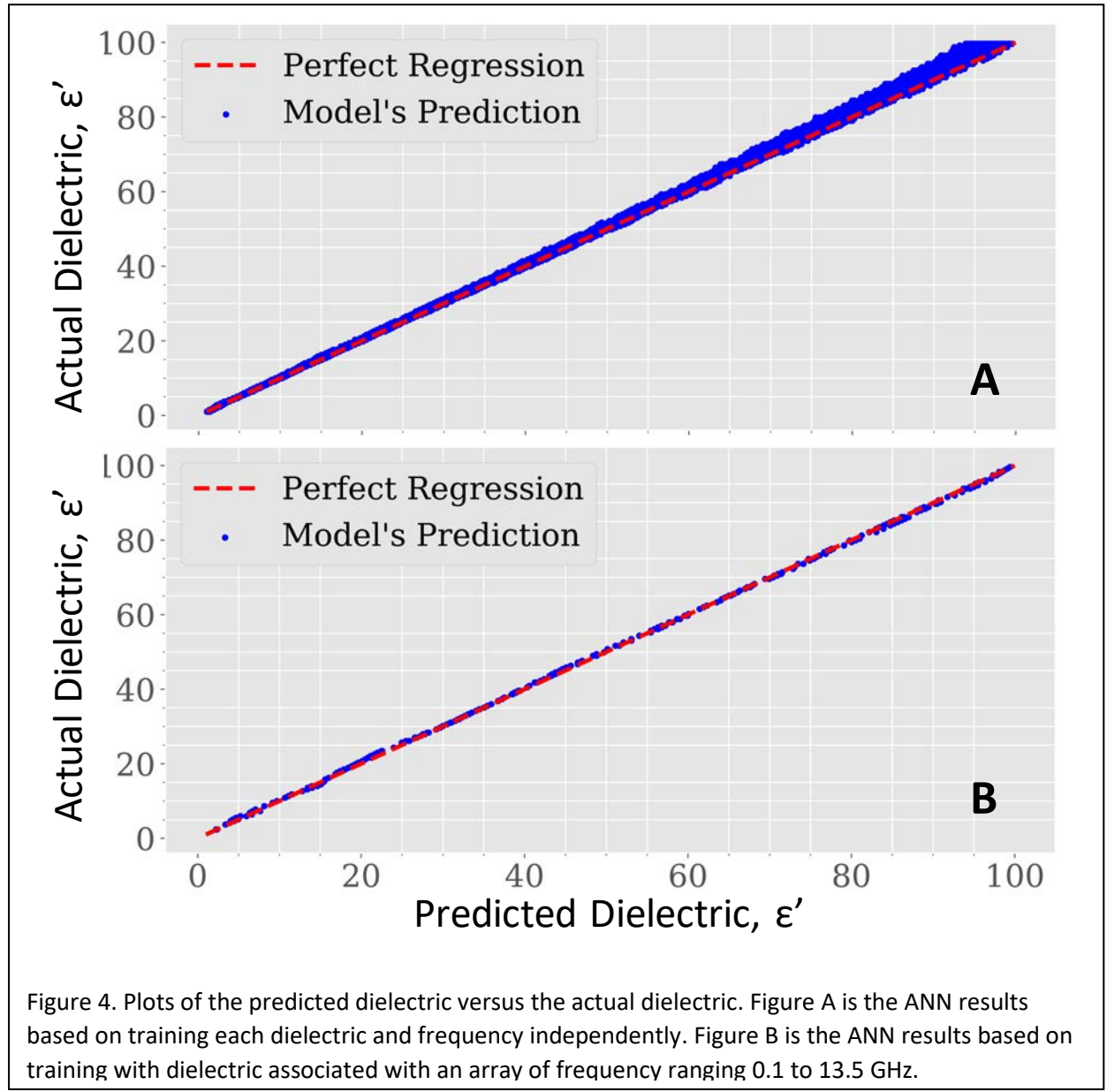

Figure 4. Plots of the predicted dielectric versus the actual dielectric. Figure A is the ANN results based on training each dielectric and frequency independently. Figure B is the ANN results based on training with dielectric associated with an array of frequency ranging 0.1 to 13.5 GHz.

Considering the first approach, Figure 4A is a plot of the results of the CNN when individual frequency points are considered. The statistical results of the network showed an MSE of 0.25 and an MAE of only 0.36, meaning that there were no outliers or parts of the frequency spectrum that the network could not learn. The network was also able to achieve an MSE and MAE of 0.001 and 0.03 respectively on the imaginary portion of the dielectric.

A comparison between Figure 4A and Figure 3A and Figure 3B confirms there is an inverse relationship between how strongly input and output features are correlated and the predictive accuracy of the ANN. As dielectric constant approaches, 100 and the correlation between the magnitudes of S11 and S21 show a much stronger correlation the predictive accuracy of the ANN goes down because of the strong dependence on the input parameters. This phenomenon is known as multicollinearity where a variable can be linearly predicted from the others with a high degree of accuracy resulting in erratic responses to small changes. The strong correlation skews the values of predictions with small changes in the weight resulting in larger responses in the output neurons. At smaller dielectrics, constants were the only strong correlation is between some bands of the phase of S11 the predictive accuracy of the ANN is much greater showing little scattering from the regression line.

The second approach results are shown in Figure 4B, where the entire frequency spectrum is considered rather than individual points. As with the results shown in Figure 4A, the network can accurately predict the dielectric constant for all values looked at in this study. However, a comparison of Subfigures 4A and 4B confirms that this second approach has a much smaller spread of predictions, especially at high dielectrics. Statistically, the results between these two approaches are very similar to this approach having an MSE of 0.43 and an MAE of 0.511 for the real portion of the dielectric. While the imaginary portion had an MSE of 0.002 and MAE of 0.035.

6.2 Experimental Data Results

To validate that the ANN architecture that was selected could be used in future applications experimentally collected data needed to be tested on it. This was accomplished using a Teflon piece of 44.45 mm in length and machined to fit the high precision coaxial airline. The validation metrics were performed on the dielectric constant of the Teflon piece as well as the other dielectric properties such as the loss tangent. The scattering parameters from the Teflon piece were evaluated using the different CNN approaches. The pre-trained ANN was loaded into python as a json file with the weights saved as an h5 file. The Teflon's scattering parameters were evaluated over the frequency range and compared to the NRW results for evaluation. The performance of ANN at predicting the dielectric constant and the loss tangent of the Teflon is shown in table one, once again the system was evaluated using the MSE and MAE. The equation for loss tangent is shown in Equation 1, epsilons are the associated components of the complex dielectric.

$$\tan(\delta) = \frac{\varepsilon''}{\varepsilon'}$$ Eq. (1)

|  | MAE Model 1 | MSE Model 1 | MAE Model 2 | MSE Model 2 |
|---|---|---|---|---|
| **ε'** | 0.24 | 0.22 | 0.56 | 0.66 |
| **tan(δ)** | 0.19 | 0.16 | 0.59 | 0.64 |

Table 2. Comparison of the predicted dielectric properties with experimentally determined dielectric properties of Teflon. The experimental values are based on NRW method with ε' = 2.16 and loss tangent = 0.0007. MSE=mean squared error and MAE=mean absolute error. Model 1 is associated with the Figure 4A and Model 2 is associated with Figure 4B.

5. Conclusions and Recommendations

A machine learning ANN can be designed that predicts the dielectric properties of any material inside of coaxial airline geometry using only the standard inputs of S11 and S21. With either method discussed in this paper showing excellent results. The system showed exceptional performance on training datasets and experimentally collected datasets. Input data required very little prepossessing, with scaling being the only numeric manipulation done to the datasets. This study shows that with a high-fidelity model of a given geometry an ANN can be created on computational data that will allow the prediction of dielectric properties without the need to de-embed air. It should be noted that as the dielectric constant increased the ANN had a harder time predicting. This problem could be eliminated using a filtering system with multiple downstream neural networks that train on smaller ranges of data to increase accuracy within ranges of interest.

As part of a larger project, the ANN developed here can help to form a vital link between in-situ reactions in the microwave regime and real-time characterization of EM wave material interactions. The methodology can be extremely helpful in characterizing things such as microwave catalysts in real-time to further the study of catalytic materials.

6. Acknowledgements

R.T and T.M. would like to acknowledge the supported in part by an appointment to the Department of Energy at National Energy Technology Laboratory, administered by ORAU through the U.S. Department of Energy Oak Ridge Institute for Science and Education. T.M. would also like to acknowledge the partial support of DE-FE0026825.

7. Data Availability

The data that support the findings of this study are available from the corresponding author upon reasonable request. The neural network model will be made available on github upon acceptance of paper.

[1]     A. Zangwill, *Modern electrodynamics*. 2013.

[2]     H. E. Bussey, "Measurement of RF Properties of Materials A Survey," *Proc. IEEE*, vol. 55, no. 6, pp. 1046–1053, 1967.

[3]     K. C. Yaw, "Measurement of dielectric material properties Application Note, Rohde & Schwarz," *Meas. Tech.*, pp. 1–35, 2006.

[4]     T. L. Blakney and W. B. Weir, "Automatic Measurement of Complex Dielectric Constant and Permeability at Microwave Frequencies," *Proc. IEEE*, vol. 63, no. 1, pp. 203–205, 1975.

[5]     A. M. Nicolson and G. F. Ross, "Measurement of the Intrinsic Properties of Materials by Time-Domain Techniques," *IEEE Trans. Instrum. Meas.*, vol. 19, no. 4, pp. 377–382, 1970.

[6]     R. G. Geyer *et al.*, "Measuring the permittivity and permeability of lossy materials," (No. Technical Note (NIST TN)-1536), 2005.

[7]     K. J. Bois, "Analysis of an Open-Ended Coaxial Probe with Lift-off for Nondestructive Testing," *IEEE Trans. Instrum. Meas.*, 1999.

[8]     J. Baker-Jarvis, M. D. Janezic, J. H. Grosvenor Jr, and R. G. Geyer, *Transmission/Reflection and Short-Circuit Line Methods for Measuring Permittivity and Permeability*. Washington D.C., 1992.

[9]     P. G. Bartley and S. B. Begley, "A new technique for the determination of the complex permittivity and permeability of materials," in *IEEE Instrumentation & Measurement Technology Concfrence*, 2010.

[10]    T. Sphicopoulos, V. Teodoridis, and F. Gardiol, "Simple Nondestructive Method for the Measurement of Material Permittivity.," *J. Microw. Power*, vol. 20, no. 3, pp. 165–172, 1985.

[11]    J. Baker-jarvis, M. D. Janezic, P. D. Domich, and R. G. Geyer, "Analysis of an Open-Ended Coaxial Probe with Lift-Off for Nondestructive Testing," vol. 43, no. 5, pp. 1–8, 1994.

[12]    M. A. Stuchly and S. S. Stuchly, "Coaxial Line Reflection Methods for Measuring Dielectric Properties of Biological Substances at Radio and Microwave Frequencies—A Review," *J. Microw. Power*, vol. 43, no. 3, pp. 165–172, 1980.

[13]    A. Nigrin, *Neural networks for pattern recognition*. MIT Press, 1993.

[14]    D. J. Scott, P. V. Coveney, J. A. Kilner, J. C. H. Rossiny, and N. M. N. Alford, "Prediction of the functional properties of ceramic materials from composition using artificial neural networks," *Journal of the European Ceramic Society*, vol. 27, no. 16. pp. 4425–4435, 2007.

[15]    L. Raff, R. Komanduri, M. Hagan, and S. Bukkapatnam, "Applications of Neural Network Fitting of Potential -Energy Surfaces," in *Neural Networks in Chemical Reaction Dynamics*, no. 2012, Oxford University Press, 2019.

[16]    Z. Z. -, Y. Q. -, Y. Z. -, and X. S. -, "Overview of Microwave Device Modeling Techniques Based on Machine Learning," *Int. J. Adv. Comput. Technol.*, vol. 5, no. 9, pp. 299–306, 2013.

[17]    D. Tuck and S. Coad, "Neurocomputed Model of Open-Circuited Coaxial Probes," *Ind. Res.*, vol. 5, no. 4, pp. 5–7, 1995.

[18]    G. Gajda and S. S. Stuchly, "An Equivalent Circuit ofan Open-Ended Coaxial Line," *IEEE Trans. Instrum. MEASIJREMENT,* vol. IM, no. 4, pp. 367–368, 1983.

[19] Q. Chen, K.-M. Huang, X. Yang, M. Luo, and H. Zhu, "AN ARTIFICIAL NERVE NETWORK REALIZATION IN THE MEASUREMENT OF MATERIAL PERMITTIVITY," *Prog. Electromagn. Res.*, vol. 116, no. April, pp. 347–361, 2011.

[20] T. L. Blakney, W. B. Weir, and D. Constant, "Comments on 'Automatic Measurement of Complex Dielectric Constant and Permeability at Microwave Frequencies,'" *Proc. IEEE*, vol. 63, no. 1, pp. 203–205, 1974.

[21] COMSOL AB, "EM Module." Stockholm, Sweden.

[22] Y. Li and Y. Yuan, "Convergence Analysis of Two-layer Neural Networks with ReLU Activation," no. Nips, pp. 1–11, 2017.

[23] P. Ramachandran, B. Zoph, and Q. V. Le, "Searching for Activation Functions," pp. 1–13, 2017.

[24] F. Agostinelli, M. Hoffman, P. Sadowski, and P. Baldi, "Learning Activation Functions to Improve Deep Neural Networks," no. 2013, pp. 1–9, 2014.

[25] A. L. Maas, A. Y. Hannun, and A. Y. Ng, "Rectifier Nonlinearities Improve Neural Network Acoustic Models," *Proceedings of the 30 th International Conference on Machine Learning*, vol. 28. p. 6, 2013.

[26] S. Narang, E. Elsen, G. Diamos, and S. Sengupta, "Exploring Sparsity in Recurrent Neural Networks," pp. 1–10, 2017.

[27] M. D. Zeiler, "ADADELTA: An Adaptive Learning Rate Method," 2012.

[28] M. Mittlböck and M. Schemper, "Explained variation for logistic regression," *Stat. Med.*, vol. 15, no. 19, pp. 1987–1997, 1996.